\documentclass{article}
\usepackage{natbib,alifeconf}

\usepackage{enumerate}
\usepackage{enumitem} 
\usepackage[utf8]{inputenc} 

\title{A Minimal Model for the Emergence of Cooperation \\
in Randomly Growing Networks}
\author{Steve Miller$^{1}$ and Joshua Knowles$^{1}$  \\
\mbox{}\\
$^{1}$School of Computer Science, University of Manchester, UK \\
stevemiller.gm@gmail.com}

\begin{document}
\maketitle

\begin{abstract}

Cooperation is observed widely in nature and is thought an essential component of many evolutionary processes, yet the mechanisms by which it arises and persists are still unclear. Among several theories, \emph{network reciprocity} --- a model of inhomogeneous social interactions --- has been proposed as an enabling mechanism to explain the emergence of cooperation. Existing evolutionary models of this mechanism have tended to focus on highly heterogeneous (scale-free) networks, hence typically assume preferential attachment mechanisms, and consequently the prerequisite that individuals have global network knowledge. Within an evolutionary game theoretic context, using the weak prisoner's dilemma as a metaphor for cooperation, we present a minimal model which describes network growth by chronological random addition of new nodes, combined with regular attrition of less fit members of the population.  Specifically our model does not require that agents have access to global information and does not assume scale-free network structure or a preferential attachment mechanism.  Further our model supports the emergence of cooperation from initially non-cooperative populations. By reducing dependency on a number of assumptions, this model offers broad applicability and as such may support an explanation of the emergence of cooperation in early evolutionary transitions, where few assumptions can be made.

\end{abstract}

\section{Introduction}
Cooperation is widespread within the natural world and considered to be important in evolutionary processes, particularly in situations where complexity increases, such as early evolutionary transitions, symbiogenesis, or the formation of multicellular organisms~\citep{smith_major_1997}. A variety of enabling mechanisms have been proposed to explain the emergence and persistence of cooperation~\citep{nowak_five_2006}, although some of these rely on certain assumptions or constraints being satisfied (e.g. familial relatedness of individuals, or the existence of higher cognitive processes). 

Among these enabling mechanisms, \emph{network reciprocity} describes how reciprocal behaviour can be promoted by the form of the connectivity between members of a population. This mechanism appears less demanding in terms of specific assumptions and so potentially offers a more general explanation: most organisms  exist within some form of network. 

A large body of research has been developed which is focused on understanding network reciprocity.  Evolutionary game theory has in particular become a common approach to such investigations, routinely with the use of the single-parameter \textit{weak} prisoner's dilemma~\citep{nowak_evolutionary_1992} as a metaphor for cooperation.  

The importance of spatial structure in explaining cooperation was first highlighted in~\citep{nowak_evolutionary_1992}.  These findings were notably developed with regards to heterogeneous networks in~\citep{santos_scale-free_2005,santos_new_2006}, where it was shown that static heterogeneous networks promote cooperation.  Heterogeneity, in network topology, refers to the range of degree values in a network, where degree $k$ represents the number of edges or connections a node may have. In a homogeneously structured network every node  has the same value of $k$. A scale-free (SF) network is considered to have high heterogeneity and has $k$ values distributed according to a power law. Scale-free networks are often assumed to be the result of preferential attachment (PA) processes (also referred to as `the rich get richer' or the `Matthew effect'~\citep{merton_matthew_1968}). A key finding in~\citep{santos_scale-free_2005} is that scale-free networks (high structural heterogeneity), result in higher levels of cooperation than random networks (low heterogenity). Given that many naturally-occurring networks are considered to have a scale-free structure and hence a power-law degree distribution~\citep{barabasi_emergence_1999,barabasi_scale-free_2003} such findings regarding network heterogeneity and cooperation have naturally generated much interest.


In a particularly interesting work~\citep{poncela_complex_2008}, cooperation in dynamic scale-free networks is demonstrated using a coevolutionary model where a preferential attachment (PA) mechanism for network growth is linked to the evolutionary success of evolving agents within a network. A typical PA system involves attachment of newcomers preferentially to those existing individuals which have more network connections.  The evolutionary preferential attachment (EPA) model differs from such an approach in that it describes a process where newcomers are more likely to attach to \emph{fitter} members of the existing population.  Prior to this work, studies of cooperation in networks had focused on the effect that the network had upon the population.  The EPA model specifically adds a causal relationship in the reverse direction, whereby agent behaviour impacts network structure, and hence also offering the possibility of a feedback mechanism.

We recently illustrated that an EPA approach, which additionally incorporates population size fluctuation, supports cooperation, specifically enabling its emergence from \emph{initially non-cooperative} founder networks~\citep{miller_population_2014}. Whilst other models within the literature have investigated pruning of networks by means of link deletion~\citep{zimmermann_cooperation_2005, santos_cooperation_2006, pacheco_active_2006, traulsen_evolutionary_2009} or to a lesser extent node deletion~\citep{ perc_evolution_2009, szolnoki_impact_2009, ichinose_robustness_2013}, our approach specifically differed from these works in that we deleted nodes on the basis of (least) fitness, thus presenting an evolutionarily representative method of population attrition.

Within this report we extend our previous findings, to develop a minimal model for cooperation in networks which shifts the evolutionary focus from the network growth mechanism to the node deletion process.   We demonstrate how such a shift reduces dependency on initial population conditions and also on scale-free network heterogeneity. In the following sections, we first revisit the two key elements of network reciprocity to illustrate how such a shift can still support cooperation; we subsequently expand on the necessity for a minimal model.

\section{Two elements of network reciprocity}

Network reciprocity is often explained with reference to heterogeneity. However the promotion of cooperation observed in the foundational work of~\citep{nowak_evolutionary_1992} is due to assortativity of agent strategies (on a homogeneous lattice structure) which results in grouping of cooperators.  

\begin{figure}[!h]
	\begin{center}
		\includegraphics[width=8.2cm]{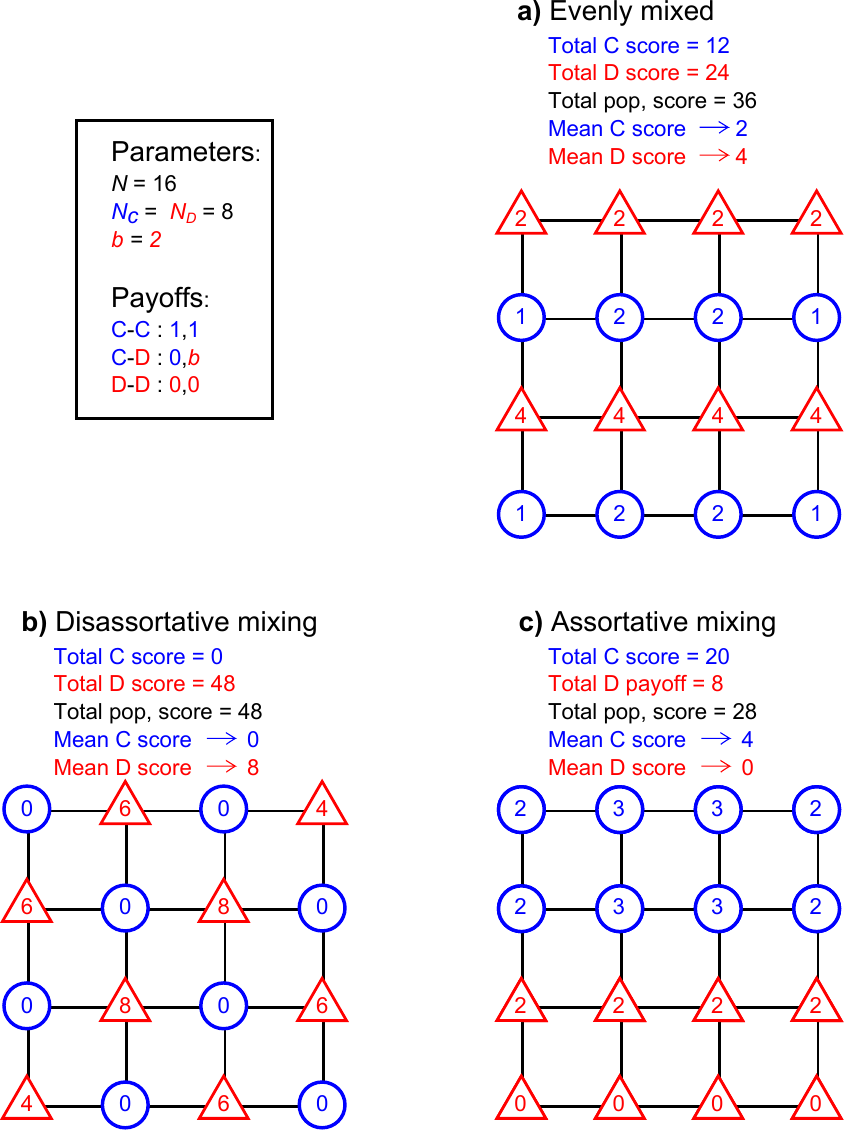}
		\caption{Sample networks illustrating how identical homogeneous network structures can support cooperation to differing extents given differing strategy distributions. Blue circles represent cooperator nodes. Red triangles represent defector nodes. Black lines represent edges (interactions) between nodes. The values within the nodes represent the scores (sum of individual edge payoffs) for the nodes. For illustrative purposes we have selected an arbitrary value of $b = 2$ to calculate example values for the scores. }
		\label{fig:three homogeneous networks}
	\end{center}
\end{figure}

Figure \ref{fig:three homogeneous networks} presents three identically structured example networks, which feature identical numbers of cooperators and defectors, in order to illustrate how assortativity can promote or suppress cooperation in the \emph{absence of heterogeneity}. These example networks can also be viewed as representative samples of larger networks with homogeneous degree distribution, in which every node has $k = 4$. We extrapolate mean scores (indicated with `$\longrightarrow$') for such larger networks which eliminate the edge effects present in the figures. 

Figure \ref{fig:three homogeneous networks}a, represents a homogeneous (evenly mixed) distribution of strategies, where each node connects to two defectors and two cooperators. Defectors outcompete cooperators in this distribution. Figure \ref{fig:three homogeneous networks}b shows a non-homogeneous (disassortative) strategy distribution where each agent connects to nonself-similar strategies.  In this case the total population score is greater than in Figure \ref{fig:three homogeneous networks}a, however this strategy distribution results in scores of zero for all cooperators, and positive scores for defectors. Figure \ref{fig:three homogeneous networks}c illustrates a non-homogeneous (assortative) strategy distribution which shows how self-similar grouping benefits cooperators but not defectors. 

From these examples, we see that, on a homogeneous spatial structure, given an ability of strategies to redistribute themselves, it is possible for cooperators to improve their lot as a result of self-assorting.  Such behaviour allows cooperators to achieve higher scores than would be achieved  for random or evenly mixed strategy distributions. From an evolutionary perspective, where scores represent fitness, it is easy to see how self-assortment may allow cooperators to outcompete defectors in a network. It should be noted that defectors do not benefit from self-assorting. (Both strategies benefit from connecting to cooperators and so defectors can therefore only benefit from `nonself-assorting' behaviour.)  

In the work of~\citep{santos_cooperation_2006} which looks at the effect of heterogeneously (rather than homogeneously) structured networks, the mechanism of assortativity illustrated above supports the formation of groups of cooperators, whilst the greater connectivity that can be found between some individuals in heterogeneous networks offers increased rewards to cooperators in groups.  Such models combine the two elements we refer to in this section: \emph{Heterogeneity} further increases the potential gains that can be made by self-\emph{assortativity}.

The volume of scientific literature on the role of heterogeneous networks in cooperation combined with the prevalence of such  networks in the real world, tend to leave strategy assortativity as a somewhat marginalised topic.  It is worth highlighting that, regardless of the heterogeneity (or lack of) in the network, cooperation cannot emerge without some form of redistribution process that supports grouping of cooperators: Assortativity remains the essential underpinning requirement for spatially structured reciprocity. In this paper we investigate a model for the emergence of cooperation in dynamic networks of evolving agents.  Our model makes few demands regarding the specifics of network structure (and associated mechanisms of network formation), whilst still promoting assortativity. We explore the rationale to produce such a model in the following section.

\section{A minimal model}

The EPA model~\citep{poncela_complex_2008} introduced earlier has been proposed as a possible explanation for the evolutionary origins of cooperation. The model uses a mechanism for network growth whereby new agents (nodes) added to a network preferentially attach to fitter nodes. The probability that an existing node \textit{i} receives one of the \textit{m} new edges is as follows:

\begin{equation}
	\Pi(t) = \frac
	{1 - \epsilon + \epsilon f_i(t)}
	{\sum_{j=1}^{N(t)}(1 - \epsilon + \epsilon f_j(t))} ,
	\label{eqn:EPA_node_addition}
\end{equation}

\noindent where $f_i(t)$ is the fitness of an existing node \textit{i} and $N(t)$ is the number of nodes available to connect to at time \textit{t} in the existing population. The parameter $\epsilon \in [0,1)$ is used to adjust selection pressure. (A fuller explanation of the details of the EPA implementation is provided in the methods section.) Inspection of Equation \ref{eqn:EPA_node_addition} highlights that in order for a newcomer to ``decide'' which node to connect to, it is required to have ``global'' information regarding i) the fitness of all other individuals in the population, and ii) the size of the population, both of which are unlikely to be satisfiable in real world examples.  

Given the preferential attachment mechanism, the EPA model is expected to generate a scale-free network (ibid.). Visual assessment of degree distribution supports this hypothesis for certain implementations i.e. for those \textit{b} values where cooperators form the majority strategy. 

The scale-free property observed in PA network models parallels many empirical findings in real networks~\citep{barabasi_scale-free_2003}, however whilst many real networks have been proposed to be scale-free (on the basis of apparent power-law degree distributions), Clauset et al. (2009) have highlighted that such claims are often hypothesised rather than demonstrated\nocite{clauset_power-law_2009}.  Complex networks are difficult to characterise with certainty (accurately distinguishing power-law distributions from e.g. stretched exponentials is a non-trivial problem) and previous claims of scale-free characteristics in real networks have subsequently been challenged~\citep{amaral_classes_2000,doyle_robust_2005,tanaka_scale-rich_2005}. A claim that a network is scale-free is plausible if a preferential attachment process is known to have generated the network, however in the absence of such knowledge, assumptions of scale-free topology may be unreliable. Network models that presuppose scale-free heterogeneity in order to explain cooperation are therefore potentially constrained by such assumptions. We also note that whilst preferential attachment models generate scale-free networks, the converse does not necessarily hold, i.e. while power-law distributions \textit{may} arise as the result of preferential attachment; other approaches can also  generate such distributions~\citep{miller_effects_1957,albert_statistical_2002, caldarelli_scale-free_2002}.

The application of simplified PA models to real world situations is also impacted by the absence of a general explanation addressing the underlying preferential attachment mechanisms. Each novel situation requires its own explanation. We consider, in particular, the question of how cooperation emerged in early evolutionary transitions. In such situations, involving primitive life forms, which might be for example, immobile, carried by currents and/or interacting randomly, it is unclear whether a mechanism may have existed by which preferential attachment occurred. The ability of fitter individuals to preferentially influence social structure (and hence drive the formation of scale-free networks) cannot be generally assumed. 

We now present a minimal model for cooperation in networks which does not require that agents have access to global knowledge, does not depend on a preferential attachment, does not require scale-free network structure, and also does not define or imply feedback between agent behaviour and the population structure.

Our model implements network growth by chronological random attachment (CRA) of new nodes alongside a strategy updating rule which defines agents' evolutionary behaviour. The network is grown by attaching new nodes randomly to existing nodes within the network. We highlight that this does not generate a typical random network (which would have a Poisson degree distribution); instead the chronological nature of the additions results in an exponential degree distribution~\citep{dorogovtsev_evolution_2002}, with `older' nodes being more highly connected. The evolutionary strategy updating, which defines strategy assortment, drives the displacement of less fit strategies by those of more successful neighbours. This updating process takes place alongside the growth of the network. For purposes of comparison, our model is based as closely as possible on the EPA model. Strategy updating rules are identical; the node attachment process differs. We also incorporate an additional mechanism not present in the original EPA model: whereas networks established by EPA are fixed in structure once the network reaches a specified size, our model incorporates an attrition process. In this process a certain proportion of less fit members of the population are removed by tournament selection whenever the network reaches a maximum size. The details of EPA and CRA models are presented in the methods section.

\section{Methods}
\noindent \textbf{Overview.} Our models and simulations are based on those described in~\citep{poncela_complex_2008}, but with the addition, in the case of attrition implementations, of a tournament selection step that removes nodes from the network. We here give a full description of the approach for completeness.

The models consist of a network (i.e. graph) with agents situated at the nodes. Edges between nodes represent interactions between agents. Interactions are behaviours between agents playing the one-shot prisoner's dilemma game. These behaviours are encoded by a `strategy' variable which takes one of two values: cooperate or defect. The game is played in a round robin fashion, with each agent playing its strategy against all its connected neighbours, in turn. Each agent thus accumulates a fitness score which is the sum of all the individual game payoffs. 

Within an evolutionary simulation, starting from a founding population, this process is repeated over generations. The evolutionary process assesses agents at each generation on the basis of their fitness score: Fitter agents' strategies remain unchanged; less fit agents are more likely to have strategies replaced by those of fitter neighbours.

The evolutionary preferential attachment (EPA) model connects strategy dynamics to network growth: starting from a small founding population, newcomer nodes are added which preferentially connect to fitter agents within the network. Our chronological random attachment (CRA) model uses the same founding population structures as EPA but adds newcomer nodes to randomly selected existing nodes. 

Attrition implementations of both models add a further component which repeatedly prunes the network: Whenever the population reaches a maximum size, a specified percentage of nodes in the network are removed, on the basis of least fitness, after which the network grows again. \\

\noindent \textbf{Outline of the evolutionary process.} Unless stated otherwise in the text, the general outline of the evolutionary process we use is described, for one generation, as follows:

\begin{enumerate}[noitemsep]
	\item  \textit{Play prisoner's dilemma}: Each agent plays one-shot prisoner's dilemma with all neighbours and achieves a fitness score that is the sum of all the payoffs.
	\item \textit{Update strategies}: Those strategies that achieve low scores are replaced on a probabilistic basis by comparison with the strategies of randomly selected neighbours.
	\item \textit{Grow network}: A specified number of new nodes are added to the network, connecting to $m$ distinct existing nodes via \textit{m} edges using either EPA or CRA. 
	\item \textit{Remove nodes (only in the case of attrition models)}: If the network has reached maximum size, it is pruned by a tournament selection process that removes less fit agents.
\end{enumerate}

\noindent In the following, we provide more detail on the specifics of each of the four steps:\\

\begin{figure}[h]
	\begin{center}
		\includegraphics[width=6cm]{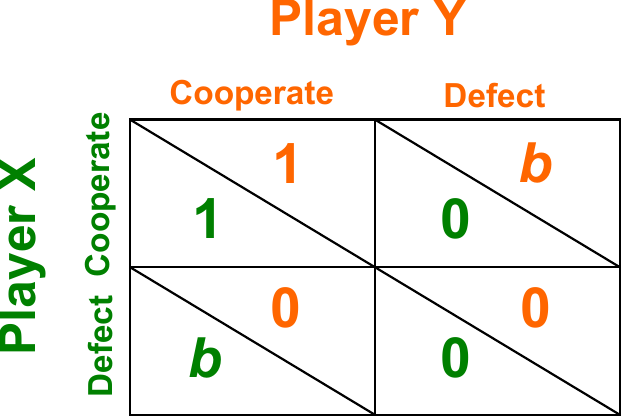}
		\caption{Payoff matrix for weak prisoner's dilemma.}
		\label{fig:PD_payoff_matrix}
	\end{center}
\end{figure}

\noindent \textbf{\textit{Play prisoner's dilemma}}. We use the single parameter representation of the one-shot prisoner's dilemma as formulated in~\citep{nowak_evolutionary_1992}. In this form (the `weak' prisoner's dilemma), payoff values for the actions, referred to as \textit{T, R, P} and \textit{S}, become $b$, 1, 0 and 0 (see Figure \ref{fig:PD_payoff_matrix}). The \textit{b} parameter represents the `temptation to defect' and is set at a value greater than 1 for the dilemma to exist. 

From the accumulated prisoner's dilemma interactions, each agent achieves a fitness score as follows:

\begin{equation}
f_i = \sum_{j=1}^{k_i} \pi_{i,j},
\label{eqn:fitness_scoring}
\end{equation}
\noindent where $k_i$ is the number of neighbours that node \textit{i} has, \textit{j} represents a connected neighbour and $\pi_{i,j}$ represents the payoff achieved by node \textit{i} from playing prisoner's dilemma with node \textit{j}.

\noindent \textit{\textbf{Update strategies}}. Each node \textit{i} selects a neighbour \textit{j} at random. If the fitness of node \textit{i}, $f_i$ is greater or equal to the neighbour's fitness $f_j$, then \textit{i}'s strategy is unchanged. If the fitness of node \textit{i}, $f_i$ is less than the neighbour's fitness, $f_j$, then \textit{i}'s strategy is replaced by a copy of the neighbour \textit{j}'s strategy, according to a probability proportional to the difference between their fitness values. Thus poor scoring nodes have their strategies displaced by the strategies of more successful neighbours. 

More precisely, at generation \textit{t}, if $f_{i}(t)\geq f_{j}(t)$ then \textit{i}'s strategy remains unchanged. If $f_{i}(t)< f_{j}(t)$ then \textit{i}'s strategy is replaced with that of the neighbour \textit{j} with the following probability:

\begin{equation}
	P_i = \frac
	{f_j(t) - f_i(t)}
	{b.max[k_i(t),k_j(t)]},
	\label{eqn:strategy_updating}
\end{equation}

\noindent where $k_{i}$ and $k_{j}$ are degrees of node \textit{i} and its neighbour \textit{j} respectively. The purpose of the denominator is to normalise the difference between the two nodes. The expression $b.max[k_{i}(t),k_{j}(t)]$ represents the largest achievable fitness difference between the two nodes given their respective degrees.

\noindent \textit{\textbf{Grow network}}. New nodes, with randomly allocated strategies, are added to achieve a total of 10 at each generation. The probability that an existing node \textit{i} receives one of the \textit{m} new edges was shown in Equation \ref{eqn:EPA_node_addition}. Each new node uses \textit{m} edges to connect to existing nodes. In all our simulations, we use \textit{m} = 2 edges. Duplicate edges and self-edges are not allowed.

Given that in our model each new node extends $m = 2$ new edges, and multiple edges are not allowed, $N$ is therefore determined \textit{without replacement}. The parameter $\epsilon \in [0,1)$ is used to adjust selection pressure. For all of our simulations $\epsilon = 0.99$, hence focusing our model on selection occurring directly as a result of the preferential attachment process.

In CRA implementations, new nodes connect to existing nodes randomly. The probability that an existing node \textit{i} receives one of the \textit{m} new edges becomes simply: 

\begin{equation}
	\Pi(t) = \frac
	{1}
	{N(t)} .
	\label{eqn:RCA_node_addition}
\end{equation}

\noindent \textit{\textbf{Remove nodes (in the case of attrition models)}}. On achieving a specified size, the network is pruned by a percentage \textit{X}. This is achieved by tournament selection using a tournament size equivalent to $1\%$ of the population. The tournament members are selected randomly from the population. The tournament member having the least fitness is the `winner'. The remaining nodes are returned to the population. By this method, a shortlist of $X\%$ nodes is established for removal from the network. All edges from deleted nodes are removed from the network. Any nodes that become disconnected from the network as a result of this process are also deleted. (Failure to do this would result in small numbers of single, disconnected, non-playing nodes, having static strategies, whose zero fitness values would result in continual isolation from the network.) When there are multiple nodes of equivalent low fitness value, the selection is effectively random (on the basis that the members were originally picked from the population randomly). Where $X = 0$, no attrition occurs.\\

\noindent \textbf{General simulation conditions}. We investigated networks grown from an initial complete network with $N_0 = 3$ agents at generation $t_0$. Founding populations were either entirely cooperators or entirely defectors. Networks were grown to a maximum size of $N = 1000$ nodes with an overall average degree of approximately $k = 4$. Simulations were run until 2000 generations. The `fraction of cooperators' values we use (denoted by $\langle{c}\rangle$) are means, averaged over the last 20 generations of each simulation, in order to compensate for variability that might occur if just using final generation values. Each simulation consisted of 10 replicates. We used $X$ = 2.5\% for all simulations.

\section{Results}

We compared the effects of non-attrition vs.\ attrition (`$+$') implementations of two models: i) chronological random attachment (CRA and CRA+), and ii) evolutionary preferential attachment (EPA and EPA+).\\

\begin{figure}[t]
	\begin{center}
		\includegraphics[width=8.2cm]{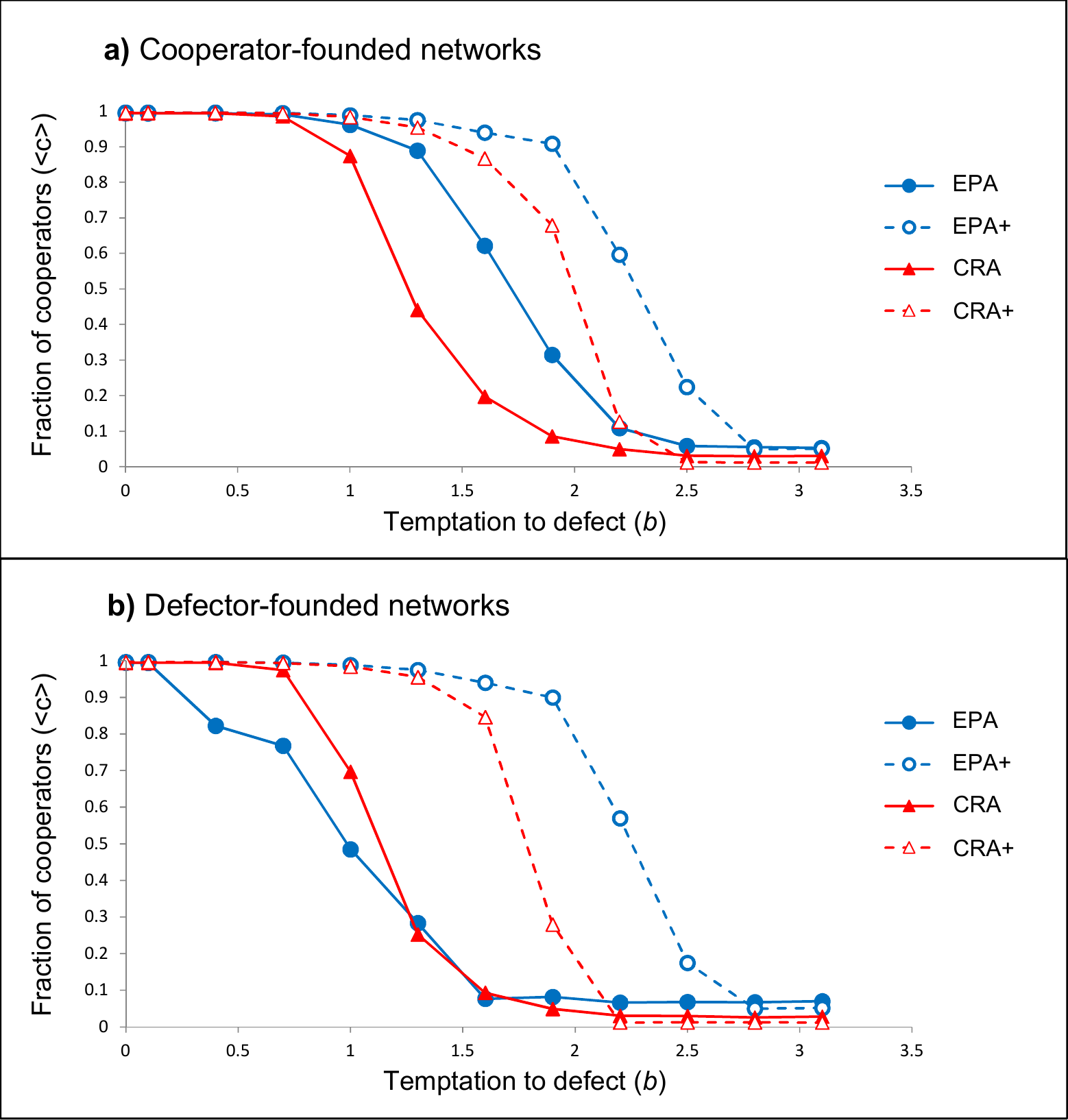}
		\caption{The effect of network model on the relationship between temptation to defect and cooperation. Each line is the average of 10 replicate simulations. Simulations featured 1000 nodes and were run for 2000 generations. Non-attrition models achieved fixed network structure after 100 generations beyond which strategy updating continued alone. Attrition implementations (EPA+ \& CRA+) are represented with dashed lines.}
		\label{fig:compare_RCA_EPA}
	\end{center}
\end{figure}

\noindent\textbf{How do preferential and random attachment affect the models?} In Figure \ref{fig:compare_RCA_EPA} we show profiles of final levels of cooperation, $\langle{c}\rangle$ vs.\ temptation to defect, \textit{b} for the four implementations. We note that whilst the preferential attachment mechanism appeared (on the basis of log-log linearity) to result in scale-free networks for those values of \textit{b} where cooperation is supported, it is hard in the light of these results to argue that such scale-free networks necessarily achieve higher levels of cooperation than those formed by random node addition (which have exponential degree distributions): Examining non-attrition implementations of the models (solid lines), we see that in the case of networks grown from cooperative founders, EPA results in higher levels of cooperation than CRA. However, the same cannot be said for populations grown from defectors where random attachment results in higher levels of cooperation for values of $b < 1.3$. Why does random attachment benefit cooperation in defector-founded networks?

\begin{figure}[!h]
	\begin{center}
		\includegraphics[width=6.55cm]{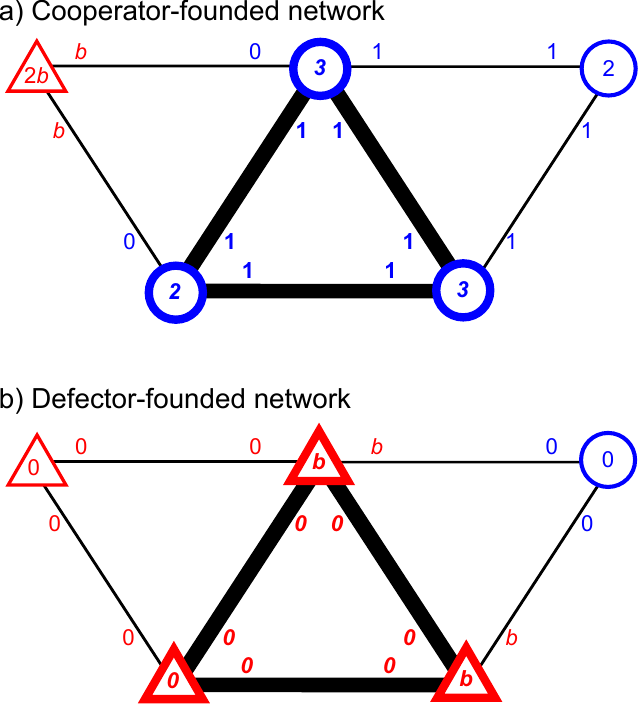}
		\caption{Initial interactions between founder networks and added nodes for \textbf{a)} cooperator- and \textbf{b)} defector-founded networks.  Blue circles represent cooperators.  Red triangles represent defectors. Interaction payoffs are shown along edges.  Cumulative scores are shown within nodes.  Founder networks are shown in bold.}
		\label{fig:early_network_effects}
	\end{center}
\end{figure}

We attempt to answer this using Figure \ref{fig:early_network_effects} where we illustrate early node attachment to cooperator- and defector-founded populations. In Figure \ref{fig:early_network_effects}a, the three cooperator founders have their scores reinforced by their interconnections (shown with bold lines). Added nodes (dashed lines) differ in their scores by a factor of \textit{b} depending on whether they are cooperator or defector.  Nodes added to a cooperator-founded network will initially tend to have scores ($= m$ or $m*b$) \textit{relatively} similar to the founder nodes ($ = 2$ or $3$), with precise values dependent on \textit{b}.  Figure \ref{fig:early_network_effects}b illustrates that self-similarity within the founder network does not benefit defectors in the same way that is seen for cooperators (D-D interactions results in payoffs of zero for both individuals). Whilst this seems to be a weakness, defector-founded networks have an alternative advantage: the addition of new nodes which are cooperators increases the founders' scores, whilst the newly added cooperators score zero. Added defectors score zero and only add to the mass of defectors present. In summary, nodes initially added to a defector-founded network score zero, regardless of strategy.

The interactions within these networks are complex and subject to random events.  Outcomes are not assured, however we see that \emph{for defector-founded networks}, relative score differences will create a high scoring founder network with newcomers unable to score.  Whereas in a cooperator-founded network, scores will be relatively similar between newcomers and existing nodes, in defector-founded networks, we see a disparity in scores which causes an initial bias against cooperation.  Cooperators in this situation are likely to be converted to defectors. Preferential attachment, by definition, drives new nodes to connect to the higher scoring founder members.  It therefore promotes the disparity in scores and reduces the likelihood of new nodes attaching to non-founders, although it does not eliminate the possibility.  (The bias against cooperation that we have described can be overcome in the less probable situation where new cooperator nodes connect to other cooperator nodes rather than the initial defector founders.)

We now refocus on the initial question of why CRA is able to support cooperation from defector-founded networks to a greater extent than EPA.  In CRA, by definition, new nodes are connected randomly.  Unlike EPA, high scoring defector-founder networks have no enhanced ability to preferentially attract newly added cooperators (and then convert them to defectors). Without this `pull' of new node connections to the defector founders, alternative interconnected groups can form, which would be more likely to support cooperation.  

We note, in terms of asymptotic outcomes for CRA networks, that given random attachment of new nodes, the greater the number of nodes in the network the smaller is the influence of the founders.  This situation is clearly very different to EPA where early nodes are likely to develop into influential hubs. Specifically for EPA simulations, in the case where defectors found a population, we see the interesting result that the ``rich get richer'' effect is detrimental to the interests of cooperators.  \\

\noindent\textbf{How does attrition change outcomes for these models?} We see from Figure \ref{fig:compare_RCA_EPA} that when the network models incorporate repeated attrition of least fit members (see dashed lines) cooperation is promoted, regardless of founder population strategy type. Attrition increases cooperation for both network formation models and reduces dependence on initial conditions. In direct comparisons, EPA with attrition achieves higher levels of cooperation than CRA with attrition. 

In the case of EPA, we know that the network structure becomes fixed  (by generation 100 in our simulations), and  any increase in cooperation thereafter is due to strategy updating.   When attrition is added to the model, cooperation increases, implying that the early fixation of network structure limits achievable levels of cooperation. The attrition model effectively removes this limit and allows the network to continually restructure simultaneously with strategy redistribution. Specifically, it has been postulated in~\citep{miller_population_2014} that random events during the earliest stages of a  network's formation can have long-term consequences for cooperation. We have illustrated one example of how such effects may arise in Figure \ref{fig:early_network_effects}. Given structural fixation, such consequences are `locked-in'.  Attrition allows for some opportunity to `unlock' the structure.  

More generally, for both CRA and EPA models, attrition targets low fitness nodes.  We can estimate some information about the type of strategies and the connectivity of such low fitness nodes.  They are likely to be: i) defectors amidst a `sea' of other defector nodes (D-D payoffs are 0, 0), or ii) marginalised defectors  with degree $k = 1$ (`terminal nodes'), connected to one other defector.  We note that low degree cooperators connected to defectors are unlikely to be evolutionarily stable: their strategies would be displaced by defection.  

It seems that attrition, by focusing on least fit members of the population, eliminates defector occupied nodes.  Replacement strategies however will then be either cooperators or defectors with equal probability, and the nodes they occupy may also reattach to more opportune positions in the network. \\

\begin{figure}[!h]
	\begin{center}
		\includegraphics[width=8.2cm]{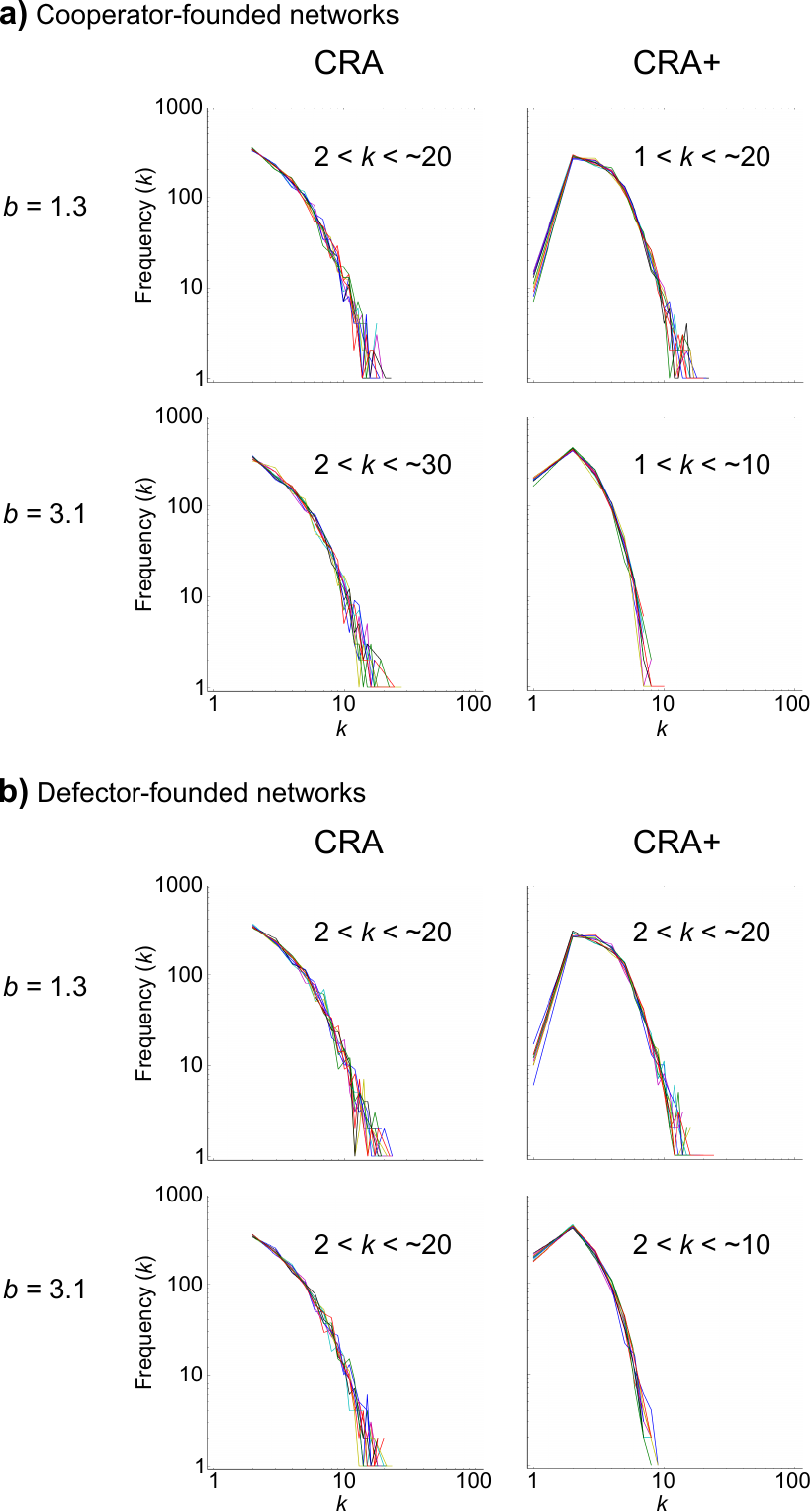}
		\caption{Degree distributions for networks formed using CRA vs.\ CRA+attrition for low and high temptation to defect (\textit{b}) values. \textbf{a}) shows results for networks grown from cooperator founders and \textbf{b}) for defector founders.  Observed range of degree values is shown within each plot.}
		\label{fig:RCA_degree_distributions}
	\end{center}
\end{figure}

\noindent\textbf{Does attrition change the topology of the CRA networks?} We show in Figure \ref{fig:RCA_degree_distributions}  that CRA+attrition does not result in a different type of degree distribution to CRA: there is no increase in the range of degree values. If anything, for very high values of \textit{b}, the opposite is true: heterogeneity of degree is reduced for the attrition networks. The reduced frequency of higher degree nodes is explained by the fact that for high values of \textit{b} the simulation will be entirely overrun by defectors so most agents will achieve scores of zero, regardless of degree. Given the uniformity of fitness values presented by this scenario, attrition becomes a random process rather than specifically being biased towards low fitness.  Note that the presence of nodes of degree $k = 1$ in the Figure is an artefact caused by the attrition process (deletion of some nodes occasionally leaves other residual nodes of degree $k = 1$ in the network).

\section{Conclusions}

We have demonstrated a model, featuring network growth by a random process, which supports network-reciprocal cooperation.  
The key to cooperation in our model is the mechanism of assortativity which allows agents in the simulation to capitalise on the exponential degree distribution of the network such that cooperators can form groups which serve to elevate the rewards available to them.  Assortativity is promoted by the evolutionary attrition process included within our model whereby less fit nodes are deleted.  

We find that our model of chronological random attachment with fitness-based attrition (CRA+), supports levels of cooperation equivalent to or greater than an existing coevolutionary method based on preferential attachment (EPA). Our model does not require that agents have memory or higher cognitive abilities and it supports cooperation, regardless of the behaviours initially present amongst the founding members of the population. Importantly, the model supports cooperation without the requirement for agents to have any form of global knowledge regarding either the network or other members of the population.  Given random linking of new nodes, there is also no requirement to explain a mechanism for preferential attachment.

Our minimal model points to a possible general explanation applicable to the emergence of cooperation in networks of primitive individuals.  The requirements are that new nodes are added over time to an existing network, and that cooperative behaviours which increase the fitness of individuals have a tendency to persist over less beneficial behaviours --- the latter being eliminated by evolutionary selection.  

We have found in comparing these models of network growth that network-reciprocal cooperation can  exist without the level of degree heterogeneity associated with scale-free structure. Such findings at first glance appear somewhat at odds with the prevailing consensus that increasing heterogeneity promotes cooperation; however by explicitly considering the benefits to cooperation offered by the combined effects of heterogeneity and assortativity, we can shed a different light on our results. Whilst our model clearly introduces only a very limited form of heterogeneity in terms of the network structure, our findings  allow for the possibility that, with regards to cooperation, it is more beneficial that cooperators can maximise a non-homogeneous self-assortative strategy distribution, than the network structure itself be highly heterogeneous.

\section{Acknowledgements}

This work has been funded by the Engineering and Physical Sciences Research Council (Grant reference number EP/I028099/1).

\footnotesize
\bibliographystyle{apalike}
\bibliography{SGMReferencesYr3}

\end{document}